\documentclass[showpacs,preprintnumbers,amsmath,amssymb,pre]{revtex4}
\usepackage{bm}
\usepackage{epsfig}
\usepackage{graphicx}
\usepackage{epstopdf} 
\usepackage[utf8]{inputenc}

\usepackage{color}

\begin{document}
\begin{large}

\title{Jamming and percolation of $k^3$-mers on simple cubic lattices}

\author{A. C. Buchini Labayen, P. M. Centres, P. M. Pasinetti, A.J. Ramirez-Pastor}

\affiliation{Departamento de F\'{\i}sica, Instituto de F\'{\i}sica Aplicada (INFAP), Universidad Nacional de San Luis - CONICET, Ej\'ercito de Los Andes 950, D5700HHW, San Luis, Argentina}

\date{\today}

\begin{abstract}

Jamming and percolation of three-dimensional (3D) $k \times k \times k $ cubic objects ($k^3$-mers) deposited on simple cubic lattices have been studied by numerical simulations complemented with finite-size scaling theory. The $k^3$-mers were irreversibly deposited into the lattice. Jamming coverage $\theta_{j,k}$ was determined for a wide range of $k$ ($2 \leq k \leq 40$). $\theta_{j,k}$ exhibits a decreasing behavior with increasing $k$, being $\theta_{j,k=\infty}=0.4204(9)$ the limit value for large $k^3$-mer sizes. In addition, a finite-size scaling analysis of the jamming transition was carried out, and the corresponding spatial correlation length critical exponent $\nu_j$ was measured, being $\nu_j \approx 3/2$. On the other hand, the obtained results for the percolation threshold $\theta_{p,k}$ showed that $\theta_{p,k}$ is an increasing function of $k$ in the range $2 \leq k \leq 16$. For $k \geq 17$, all jammed configurations are non-percolating states, and consequently, the percolation phase transition disappears. The interplay between the percolation and the jamming effects is responsible for the existence of a maximum value of $k$ (in this case, $k = 16$) from which the percolation phase transition no longer occurs. Finally, a complete analysis of critical exponents and universality has been done, showing that the percolation phase transition involved in the system has the same universality class as the 3D random percolation, regardless of the size $k$ considered.

\end{abstract}


\maketitle

\noindent $\dag$ To whom all correspondence should be addressed.

\newpage

\section{Introduction}


Percolation theory and cluster concepts have been extremely useful in elucidating many problems in physics \cite{Stauffer,Sahimi,Grimmett,Sornette,Bollobas}. In most cases the theory predicts a geometrical transition at the percolation threshold, characterized in the percolative phase by the presence of a giant cluster, which becomes infinite in the thermodynamic limit.

One of the applications of percolation theory is connected to the study of physical and chemical properties of adsorbed monolayers. In this framework, deposition of extended objects on different surfaces is of considerable interest for a wide range of applications in biology, nanotechnology, device physics, physical chemistry, and materials science. Theoretically, several models have been developed to capture the basic physics of this situation, and by far the most studied is that of random sequential adsorption.

The random sequential adsorption (RSA), introduced by Feder \cite{Feder}, has served as a paradigm for modeling irreversible deposition processes. The main features of the RSA model are \cite{Evans}: (1) the objects are put on randomly chosen sites, (2) the adsorption is irreversible and (3) at any time only one object is being adsorbed, so that the process takes place sequentially. The final state generated by RSA is a disordered state (known as jamming state), in which no more objects can be deposited due to the absence of free space of appropriate size and shape. The jamming state has infinite memory of the process and the orientational order is purely local. In addition, the limiting or jamming coverage strongly depends on the shape and size of the depositing particles.


As it was mentioned, the percolation transition is based on calculating the probability of occurrence of an infinite connectivity among the elements occupying on the lattice. Thus, the jamming coverage has an important role on the percolation threshold and the interplay between jamming and percolation has been discussed in several works \cite{Evans,Budi1,Budi2,Redner,Stauffer,Becklehimer,Vandewalle,Corne1,Corne2,Leroyer,Bonnier,Kondrat,
Tara2012,Kondrat2017,Slutskii,EPJB4,PHYSA38,Nakamura86,Nakamura87,Centres2018,Baldosas3D}. In the following, we will discuss the main results obtained on one-dimensional \cite{Stauffer,Redner,Evans} and square plaquette lattices \cite{Becklehimer,Vandewalle,Corne1,Corne2,Leroyer,Bonnier,Kondrat,Tara2012,Kondrat2017,Slutskii,EPJB4,PHYSA38,Nakamura86,Nakamura87,Centres2018,Baldosas3D}, which is the topic of this paper.

In the case of straight rigid $k$-mers on one-dimensional (1D) lattices, the RSA problem has been exactly solved and an explicit expression for $\theta(t)$ has been derived \cite{Redner}. From the equation of $\theta(t)$, the dependence on $k$ of the jamming coverage $\theta_{j,k}$ can be obtained. Note that $\theta(t)$ represents the fraction of lattice sites covered at time $t$ by the deposited objects and, consequently, $\theta(t=\infty)=\theta_{j,k}$. For $k \rightarrow \infty$, the jamming threshold tends to R\'enyi's Parking constant for the continuous case $\theta_{j,k} \rightarrow c_R \approx 0.7475979202$ \cite{Renyi}. On the other hand, the percolation problem of linear $k$-mers on 1D lattices is trivial: the percolating cluster appears only for $k=1$ and full coverage ($\theta=1$) \cite{Stauffer,Evans}. For $k>1$, the jamming coverage is less than 1 and consequently, the percolation phase transition disappears.


Several authors investigated the isotropic deposition of straight rigid $k$-mers on two-dimensional (2D) square lattices \cite{Becklehimer,Vandewalle,Corne1,Corne2,Leroyer,Bonnier,Kondrat,Tara2012,Kondrat2017,Slutskii}. In Ref.\cite{Becklehimer}, linear $k$-mers with a length in the interval $k = 1, . . . , 20$ were randomly and isotropically deposited on a 2D square lattice. By computer simulations, the authors found that the percolation threshold decreases with increasing the chain length $k$. A similar behavior was observed by Vandewalle et al. \cite{Vandewalle} for sizes $k$ ranging from 1 to 10, and by Cornette et al. \cite{Corne1,Corne2} for sizes $k$ ranging from 1 to 15.

Later papers extended the analysis to larger lattices and longer objects \cite{Leroyer,Bonnier,Kondrat}. The results obtained revealed that: (1) the jamming concentration monotonically decreases and tends to $0.660(2)$ as the length of the rods increases; (2) the percolation threshold is a nonmonotonic function of the size $k$: it decreases for small rod sizes, goes through a minimum around $k=13$, and finally increases for large segments; and (3) the ratio of the two thresholds $\theta_{p,k}/\theta_{j,k}$ has a more complex behavior: after initial growth, it stabilizes between $k=3$ and $k=7$, and then it grows again.

Tarasevich et al. \cite{Tara2012} confirmed that the percolation threshold initially decreases, passes through a minimum at $k=13$, and then increases with increasing $k$. In addition, the authors determined that the percolation phase transition only exists for values of $k$ between 1 and approximately $1.2 \times 10^4$. For larger $k$, percolation cannot occur even at the jamming concentration, which is the maximum the system can achieve. Very recently, Kondrat et al. \cite{Kondrat2017} refuted the conjecture that in the RSA processes of linear $k$-mers on square lattices the percolation is impossible if the needles are sufficiently long \cite{Tara2012}. The authors presented a strict proof that in any jammed configuration of nonoverlapping, fixed-length, horizontal or vertical needles on a square lattice, all clusters are percolating clusters. The theoretical result obtained in Ref. \cite{Kondrat2017} was recently corroborated using simulation techniques \cite{Slutskii}. Based in a very efficient parallel algorithm, Slutskii {\it et al.} \cite{Slutskii} studied the problem of large linear $k$-mers (up to $k=2^{17}$) on a square lattice with periodic boundary conditions. The obtained results indicate that, in the case of linear $k$-mers on square lattices, percolation always occurs before jamming.


In Ref. \cite{EPJB4}, the percolation problem corresponding to linear $k$-mers was extended to three-dimensional (3D) simple cubic lattices. The $k$-mers were irreversibly and isotropically deposited into the lattice. Then, the percolation threshold and critical exponents were obtained by numerical simulations and finite-size scaling theory. The results, obtained for $k$ ranging from 1 to 100, revealed that (i) the percolation threshold exhibits a decreasing function when it is plotted as a function of the $k$-mer size; and (ii) the phase transition occurring in the system belongs to the standard 3D percolation universality class regardless of the value of $k$ considered.

Later, the deposition kinetics of linear $k$-mers on cubic lattices was investigated in Ref. \cite{PHYSA38}. The study revealed that (i) the jamming coverage exhibits a decreasing function when it is plotted in terms of the $k$-mer size, being $0.4045(19)$ the value of the limit coverage for large $k$'s; and(ii) the ratio between percolation threshold and jamming coverage shows a non-universal behavior, monotonically decreasing to zero with increasing $k$. These findings indicate that the percolation phase transition occurs for all values of $k$.


The RSA problem becomes more difficult to solve when the depositing particles are compact objects, and only very moderate progress has been reported so far. In the line of present work, M. Nakamura \cite{Nakamura86,Nakamura87} studied the problem of $k \times k$ square tiles ($k^2$-mers) irreversibly deposited on 2D square lattices. The author showed that the percolation threshold is an increasing function of $k$ in the range $1 \leq k \leq 3$. For $k \geq 4$, all jammed configurations are non-percolating states, and consequently, the percolation phase transition disappears. This finding was corroborated by theoretical analysis based on exact calculations of all the possible configurations on finite cells \cite{Centres2018}.


Jamming and percolation properties change substantially when the $k^2$-mers are deposited on 3D simple cubic lattices. This problem was investigated in Ref. \cite{Baldosas3D}. The jamming coverage was found to decrease to a nonzero constant with $k$ as the power law $A+B/k+C/k^2$ obtaining the fitting value $A=\theta_{j,k\rightarrow\infty}=0.4285(6)$. On the other hand, a nonmonotonic size dependence was observed for the percolation threshold, which decreases for small particles sizes, goes through a minimum around $k=18$, and finally increases for large segments. As in the case of linear $k$-mers deposited on 2D square lattices, it would be expected that the percolation phase transition survives as $k \rightarrow \infty$.


In this paper, we have studied jamming and percolation aspects of $k \times k \times k $ cubic objects ($k^3$-mers) deposited on 3D simple cubic lattices. Using extensive simulations supplemented by finite-size scaling analysis, jamming coverages and percolation thresholds were determined for a wide range of $k$ values. This study (i) completes previous work on jamming and percolation of extended objects on $D$-dimensional lattices \cite{Becklehimer,Vandewalle,Corne1,Corne2,Leroyer,Bonnier,Kondrat,Tara2012,Kondrat2017,Slutskii,EPJB4,PHYSA38,Nakamura86,Nakamura87,Centres2018,Baldosas3D}; and (ii) allow us to extract general conclusions about the behavior of the system and its dependence on the relationship between the dimension of the depositing object and the dimension of the substrate.


The paper is organized as it follows: the model is presented in Section \ref{modelo}. Jamming and percolation properties are studied in Sections \ref{jam} and \ref{perco}, respectively. Finally, the conclusions are drawn in Section \ref{conclu}.

\section{Model and Monte Carlo Simulation Details} \label{modelo}

We consider a substrate composed by a 3D simple cubic lattice of $M=L \times L \times L$ sites with periodic boundary conditions in each direction (a torus). In this way, all the lattice sites are equivalent and there are no edge effects in the filling process. The deposition procedure is as follow. A set of $k \times k \times k$ nearest-neighbor sites, conforming a cube of side $k$ (see Fig. 1), is randomly chosen; if all the selected sites are empty, a $k^3$-mer is deposited onto the lattice (the $k^3$ sites are marked as occupied). Otherwise, the attempt is rejected. When $N$ $k^3$-mers are deposited, the concentration is $\theta=k^3N/M$. In order to efficiently occupy the lattice sites, we actually select at random the empty $k^3$-tuples from a list of empty $k^3$-tuples, instead of from the whole lattice, updating the list in each step. This strategy, although it is more memory consuming, significantly improves the computational cost of the filling algorithm.

\begin{figure}[h!]
	\centering
	\includegraphics[width=10cm,clip=true]{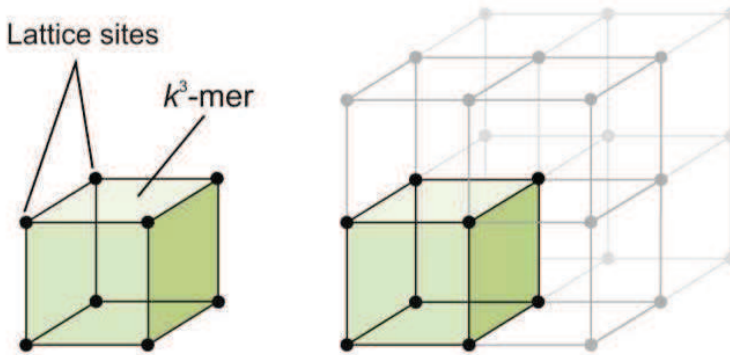}
	\caption{Schematic diagram of the system for the case $k=2$.}
	\label{fig1}
\end{figure}

\section{Jamming} \label{jam}

The calculation of the jamming concentration for different values of $k$ on a lattice of linear size $L$ (a $L$-lattice) is carried out by using the probability $W_{L,k}(\theta)$ that a particular RSA process reaches the coverage $\theta$ \cite{PHYSA38}. The procedure to determine this probability consists of simulating the following steps: (a) the setup of an initially empty cubic $L$-lattice, and (b) the deposition of the objects on the lattice until reaching a jamming condition. $n$ runs of the steps (a)-(b) are carried out for each lattice size $L$ and each object size $k$. Then the probability $W_{L,k}(\theta) = n_{L,k}(\theta)/n$ can be calculated, where $n_{L,k}(\theta)$ is the number of runs that reach the coverage $\theta$. A set of $n=10^5$ independent runs was numerically obtained for several values of the lattice sizes $L$ as well as for different values of $k$. The $L/k$ ratio is kept constant to prevent any spurious effects.

For an infinite system ($L \rightarrow \infty$) $W_{L,k}(\theta)$ should be a step function, being 1 for $\theta \leq \theta_{j,k}$ and 0 for $\theta > \theta_{j,k}$, whereas for finite values of $L$, $W_{L,k}(\theta)$ varies continuously between 1 and 0, with a sharp fall around $\theta_{j,k}$. Thereby, the jamming coverage can be estimated from the curves of $W_{L,k}$ versus $\theta$ plotted for several lattice sizes \cite{PHYSA38}. In the vicinity of the limit coverage, the probabilities show a strong dependence on the system size. However, at the jamming point, the probabilities adopt an unique value $W^*_{L,k}$, irrespective of the system sizes in the scaling limit. Thus, plotting $W_{L,k}(\theta)$ for different linear sizes $L$ yields an intersection point $W^*_{L,k}$, which gives an accurate estimation of the jamming coverage, $\theta_{j,k}$, in the infinite system. The interval width where the curves cross each other is taken as the error in the determination of $\theta_{j,k}$.

In Fig. 2, the probabilities $W_{L,k}(\theta)$ are shown for values of $L/k$ ranging from 8 to 96, as indicated, and three typical cases: $k=2$, $k=4$ and $k=12$. From the inspection of the figure (and from data not shown here for a sake of clarity), it can be seen that: (a) for each $k$, the curves cross each other in a unique point $W^*_{L,k}$; (b) those points do not modify their numerical value for the different cases studied, being $W^*_{L,k} \approx 0.49$; (c) those points are located at very well defined values in the $\theta$-axes determining the jamming threshold, $\theta_{j,k}$, for each $k$; and (d) $\theta_{j,k}$ decreases for increasing values of $k$.

\begin{figure}[h!]
	\centering
	\includegraphics[width=12cm,clip=true]{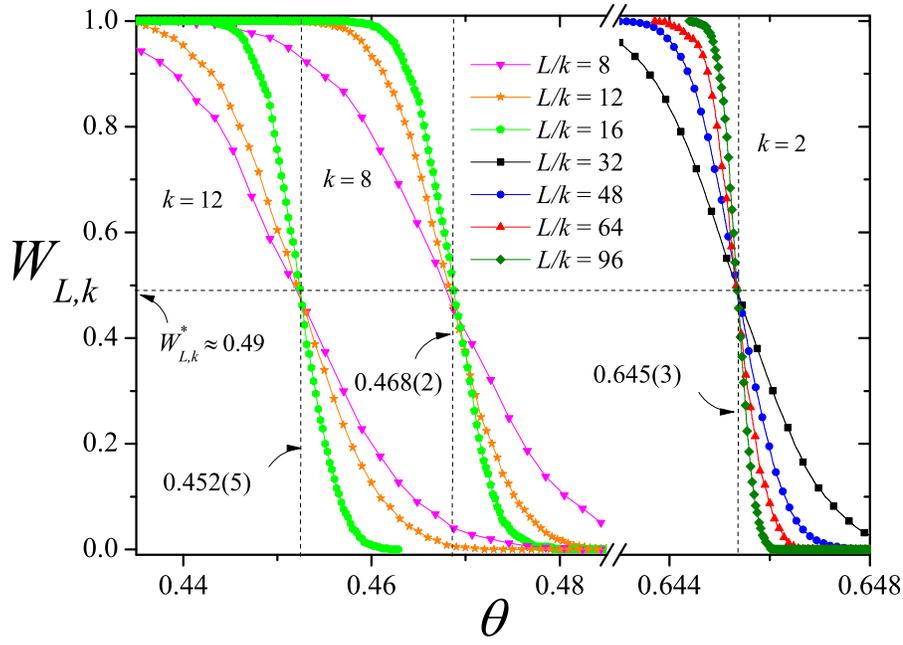}
	\caption{Curves of $W_{L,k}$ as a function of the density $\theta$ for three values of $k$-mer size (from right to left, $k=2$, $k=8$, and $k=12$) and lattice sizes ranging between $L/k=8$ and $L/k=96$ as indicated. Horizontal dashed line shows the $W_{L,k}^*$ point. Vertical dashed lines denote the jamming thresholds in the thermodynamic limit.}
	\label{fig2}
\end{figure}

\begin{figure}[h!]
	\centering
	\includegraphics[width=12cm,clip=true]{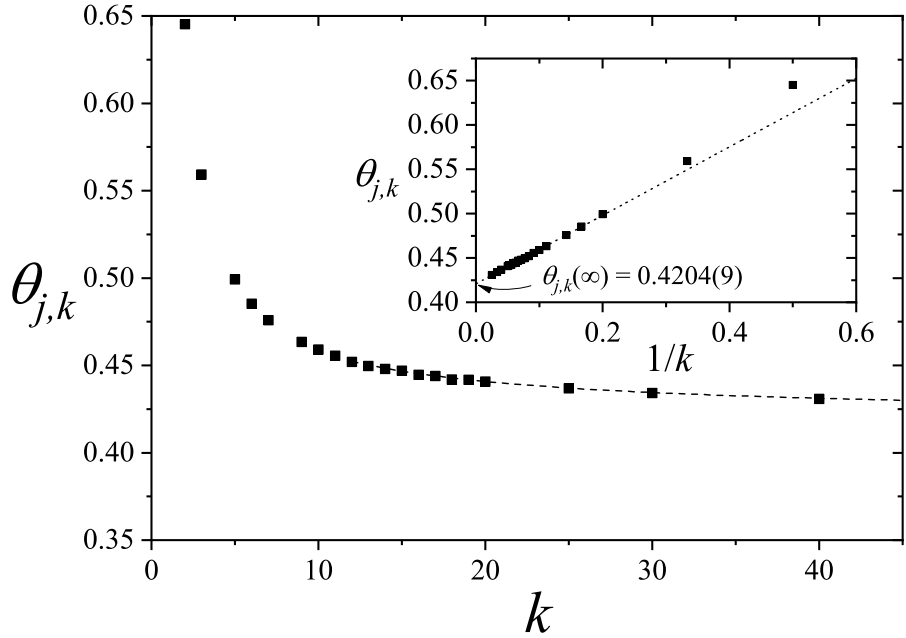}
	\caption{Jamming coverage $\theta_{j,k}$ as a function of $k$. Symbols represent simulation results and dashed line corresponds to the fitting function as discussed in the text. Inset: $\theta_{j,k}$ as a function of $1/k$.}
	\label{fig3}
\end{figure}

\begin{table}
\label{T1}
\begin{center}
\caption{Table I: Jamming and Percolation thresholds versus $k$.}
\begin{tabular}{p{1cm} |p{2.5cm} |p{2.5cm}}
$k$ &$\theta_{j,k}$ &$\theta_{p,k}$  \\
\hline

2	& 0.645(4) & 0.259(2) \\
3	& 0.559(1) & 0.268(6) \\
4	& 0.521(1) & 0.291(3) \\
5	& 0.499(2) & 0.312(4) \\
6	& 0.485(2) & 0.332(5) \\
7	& 0.476(2) & 0.352(3) \\
8	& 0.468(3) & 0.367(7) \\
9	& 0.463(3) & 0.382(1) \\
10	& 0.459(4) & 0.397(3) \\
11	& 0.456(3) & 0.408(3) \\
12	& 0.452(5) & 0.418(3) \\
13	& 0.451(5) & 0.426(1) \\
14	& 0.447(5) & 0.432(8) \\
15	& 0.447(5) & 0.438(2) \\
16	& 0.447(5) & 0.440(7) \\
17	& 0.445(4) & - \\
18	& 0.442(8) & - \\
19	& 0.442(8) & - \\
20	& 0.441(7) & - \\
25	& 0.436(5) & - \\
30	& 0.434(5) & - \\
40	& 0.430(5) & - \\

\end{tabular}
\end{center}
\end{table}

The procedure of Fig. \ref{fig2} was repeated for $k$ ranging between 2 and 40. The results are shown in Fig. \ref{fig3} and compiled in the second column of Table I. A decreasing behavior is observed for $\theta_{j,k}$, with a finite value of saturation in the limit of infinitely long $k^3$-mers. The simulation data have been fitted to the function: $\theta_{j,k}= A + B/k + C/k^2$ $(k \geq 12)$, being $A=\theta_{j,k=\infty}$=0.4204(9), $B$=0.44(3) and $C$=-0.75(30). The fitting curve is shown in Fig. \ref{fig3} (dashed line). For $k \geq 5$, the inset of Fig. 3 shows a practically linear dependence of $\theta_{j,k}$ as a function of $1/k$, which allows us to corroborate the previously obtained value $\theta_{j,k \rightarrow \infty}=0.4204(9)$.

It is interesting to compare the results in Fig. \ref{fig3} with similar data resulting from the deposition of linear $k$-mers (objects of $k \times 1 \times 1$) \cite{PHYSA38} and tiles or $k^2$-mers (objects of $k \times k \times 1$) \cite{Baldosas3D} on simple cubic lattices. The main similarities and differences are as follows: $(1)$ in the three cases, the jamming coverage was also found to be a decreasing function of $k$; $(2)$ in the range studied by simulations ($2 \leq k \leq 40$), the $k^3$-mers are less effective in filling the 3D cubic lattice than other less compact objects. As an illustrative example, $\theta_{j,k=20} \approx $ 0.5256, 0.4820 and 0.4407, for $k$-mers, $k^2$-mers and $k^3$-mers, respectively; and $(3)$ the tendency described in point $(2)$ does not seem to be valid for large values of $k$. Thus, $\theta_{j,k=\infty}$=0.4045(19) \cite{PHYSA38}, 0.4285(6) \cite{Baldosas3D} and 0.4204(9), for $k$-mers, $k^2$-mers and $k^3$-mers, respectively. The limiting values of $\theta_{j,k}$ were obtained by simulations for relatively small $k$ sizes and then extrapolated to represent very long objects. Additional simulation research of RSA with extremely long objects should be performed in the future to confirm or reject the prediction in point $(3)$.

\begin{figure}[h!]
	\centering
	\includegraphics[width=12cm,clip=true]{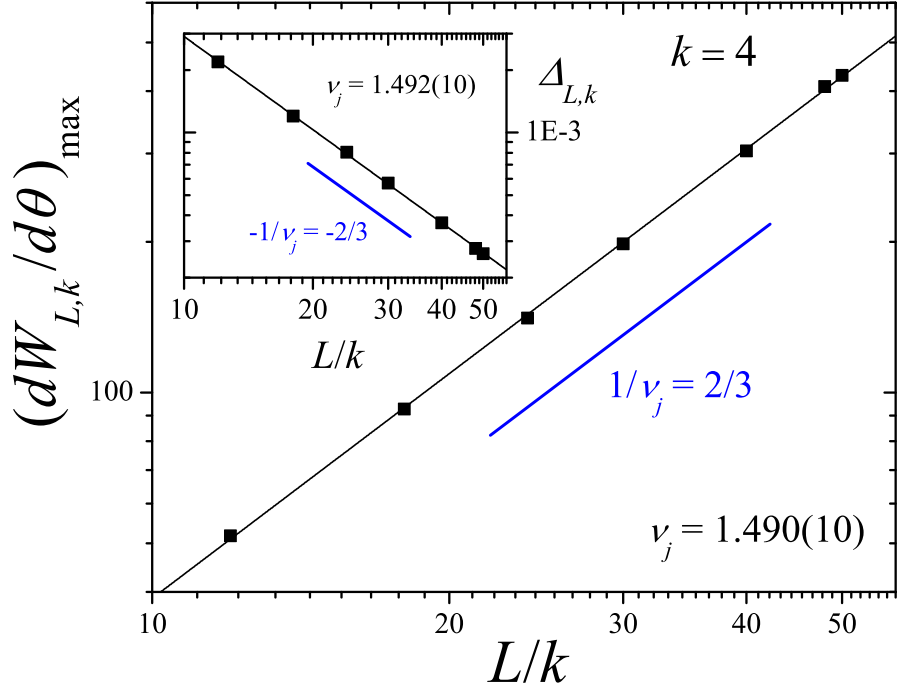}
	\caption{Log-log plot of $(dW_{L,k}/d\theta)_{\rm max}$ as a function of $L/k$ for $k$=4. According to Eq. (\ref{nuj1}) the slope of the line corresponds to 1/$\nu_{j}$. Inset: Log-log plot of the standard deviation $\Delta_{L,k}$ in Eq. (\ref{gauss}) as a function of $L/k$ for the same case shown in the main part of this figure. According to Eq. (\ref{nuj2}), the slope of the line corresponds to -1/$\nu_{j}$.}
	\label{fig4}
\end{figure}

To further deepen the study of jamming, the critical exponent $\nu_j$ of the jamming transition was calculated. For this purpose, the derivative $dW_{L,k}/d\theta$ is expected to behave like a Gaussian function around the maximum \cite{Vandewalle},
\begin{equation}\label{gauss}
{{d{W_{L,k}}} \over {d\theta }} = {1 \over {\sqrt {2\pi } {\Delta _{L,k}}}}\exp \left\{ { - {1 \over 2}{{\left[ {{{\theta  - {\theta _{j,k}}(L)} \over {{\Delta _{L,k}}}}} \right]}^2}} \right\},
\end{equation}
where $\theta_{j,k}(L)$ is the concentration at which the slope of $W_{L,k}$ is maximum and $\Delta_{L,k}$ is the standard deviation from $\theta_{j,k}(L)$. Then, by fitting the derivative according to Eq. (\ref{gauss}) it is possible to determine $\nu_j$:
\begin{equation}\label{nuj1}
{\left( {{{d{W_{L,k}}} \over {d\theta }}} \right)_{\max }} \propto {L^{1/{\nu _j}}}.
\end{equation}

Fig. \ref{fig4} shows $(dW_{L,k}/d\theta)_{max}$ as a function of $L/k$  for $k=4$ in a log-log graph. $\nu_j$ can be obtained from the slope of the curve (the line is a linear fit of the points). In this case, $\nu_j= 1.490(10)$.

An alternative way to obtain $\nu_j$ is from the divergence of the jamming standard deviation at the critical point,
\begin{equation}\label{nuj2}
\Delta_{L,k} \propto {L^{-1/{\nu _j}}}.
\end{equation}
The inset in Fig. \ref{fig4} shows $\Delta_{L,k}$ as a function of $L/k$ for the same case of the main figure. With this scheme, the resulting value of the critical exponent was $\nu_j= 1.492(10)$. In both cases (main figure and inset), the critical exponent obtained from the slope of the curves is close to $3/2$. The procedure was repeated for different values of $k$. In all the cases, the values obtained for $\nu_j$: (1) remain close to $3/2$, and (2) coincide, within the numerical errors, with the values previously reported by us in other 3D systems \cite{PHYSA38,Baldosas3D}.

\section{Percolation} \label{perco}

As it was already mentioned, the main goal of percolation theory is the determination of the minimum concentration $\theta=\theta_{p,k}$ for which a cluster extends from one side of the system to the opposite. We are interested in determining: (i) the dependence of $\theta_{p,k}$ as a function of the size $k$, and (ii) the critical exponents and the universality class of the phase transition occurring in the system.

The finite-scaling theory gives us the basis to determine the percolation threshold and the critical exponents of a system with a reasonable accuracy. For this purpose, the probability $R=R^X_{L,k}(\theta)$ that an $L$-lattice percolates at the concentration $\theta$ of occupied sites by cubic objects of size $k \times k \times k$ can be defined
\cite{Stauffer,Binder,Yone1}. Here, the following criteria can be given according to the meaning of $X$:
\begin{itemize}
  \item $R^{x}_{L,k}(\theta)$: the probability of finding a percolating cluster along the $x$-direction,
  \item $R^{y}_{L,k}(\theta)$: the probability of finding a percolating cluster along the $y$-direction,
  \item $R^{z}_{L,k}(\theta)$: the probability of finding a percolating cluster along the $z$-direction.
\end{itemize}
Other useful definitions for the finite-size analysis are:
\begin{itemize}
  \item $R^{U}_{L,k}(\theta)$: the probability of finding a cluster which percolates on any direction,
  \item $R^{I}_{L,k}(\theta)$: the probability of finding a cluster which simultaneously percolates in the three $(x,y,z)$ directions,
  \item $R^{A}_{L,k}(\theta)$=$\frac{1}{3}[R^{x}_{L,k}(\theta)+R^{y}_{L,k}(\theta)+R^{z}_{L,k}(\theta)]$: the arithmetic average.
\end{itemize}

Computational simulations were applied to determine each of the previously mentioned quantities. Each simulation run consists of the following steps: (a) the construction of a simple cubic lattice of linear size $L$ and coverage $\theta$, (b) the cluster identification using the union-and-find algorithm \cite{Hoshen} with open boundary conditions. In the last step, the size of the largest cluster $S_L$ is determined, as well as the existence of a percolating island and all the probabilities $R^{X}_{L,k}$.

A total of $m_{L}$ independent runs of such two steps procedure were carried out for each lattice size $L$. From these runs, a number $m^X_{L,k}$ of them present a percolating cluster according to the criterion $X = {x, y, z, I, U, A}$. Then, $R^X_{L,k}(\theta)= m^X_{L,k} / m_{L}$ is defined and the procedure is repeated for different values of $L$, $\theta$ and $k$.

In addition to the different probabilities  $R^X_{L,k}(\theta)$, the percolation order parameter $P$ and the corresponding susceptibility $\chi$ have been measured \cite{Biswas,Chandra},
\begin{equation}\label{parord}
P=\langle S_{L}\rangle/M,
\end{equation}
and
\begin{equation}\label{chi}
\chi=[\langle S_{L} ^2\rangle-\langle
S_{L}\rangle ^2]/M,
\end{equation}
where $\langle ... \rangle$ means an average over simulation runs.

In our percolation simulations, we used $m_{L}= 10^5$. In addition, for each value of $\theta$, the effect of finite size was investigated by examining cubic lattices with $L/k = 6, 8, 10, 12, 15$ and $24$. As it can be appreciated, this represents extensive calculations from the numeric point of view (with an effort reaching almost the limits of our computational capabilities). From there on, the finite-scaling theory can be used to determine the percolation threshold and the critical exponents with a reasonable accuracy.

An initial way to estimate the percolation threshold \cite{Yone1} is from the interception of the curves of $R^X_{L,k}(\theta)$. To improve the accuracy, different curves are expressed as a function of continuous values of $\theta$. Then, as in the case of jamming probability, $dR^R_{L,k}/d\theta$ can be fitted by the Gaussian function \footnote{The behavior of $dR^{X}_{L,k}(p)/dp$ is known not to be a Gaussian in all range of coverage \cite{Newman3}. However, this quantity is approximately Gaussian near the peak, and fitting with a Gaussian function is a good approximation for the purpose of locating its maximum.},
\begin{equation}\label{gauss2}
{{d{R^X_{L,k}}} \over {d\theta }} = {1 \over {\sqrt {2\pi } {\Delta^X _{L,k}}}}\exp \left\{ { - {1 \over 2}{{\left[ {{{\theta  - {\theta^X _{p,k}}(L)} \over {{\Delta^X _{L,k}}}}} \right]}^2}} \right\},
\end{equation}
where $\theta^X_{p,k}(L)$ and $\Delta^X_{L,k}$ have the same meaning as in Eq. (1).

\begin{figure}[h!]
	\centering
	\includegraphics[width=7cm,clip=true]{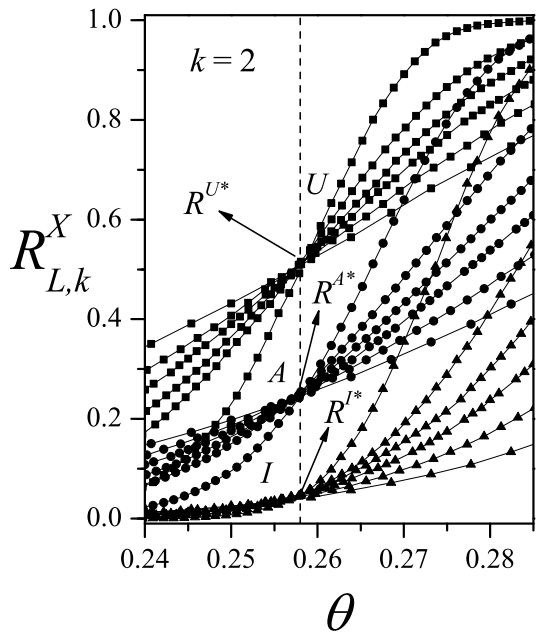}
	\caption{Fraction of percolation lattices as a function of the concentration $\theta$ for $k=2$ and several lattice sizes as indicated. The different groups of curves correspond to the criteria $U$, $A$, and $I$, from up to down.}
	\label{fig5}
\end{figure}

The probability $R^{X}_{L,k}(\theta)$, which represents the percolation cumulant and whose properties are identical to those of the Binder cumulant $U_L$ in standard thermal transitions \cite{Binder,Privman}, obeys the same scaling relation as $U_L$, and the intersection of the curves of $R^{X}_{L,k}(\theta)$ for different system sizes can be used to determine the critical point that characterizes the phase transition occurring in the system \cite{Stauffer,Corne2,Newman,Fortunato1,Fortunato2}. This procedure is shown in Fig. \ref{fig5}, where the probabilities $R^{U}_{L,k}(\theta)$, $R^{I}_{L,k}(\theta)$ and $R^{A}_{L,k}(\theta)$ are shown for $k=2$ and different lattice sizes as indicated. From a simple inspection of the figure (and from data not shown here for the sake of clarity) it is observed that: (a) the curves, corresponding to the various percolation criteria ($U$, $A$, $I$, etc.), cross each other in a unique universal point, $R^{X*}$, which depends on the criterion $X$ used; and (b) those points are located at well defined values in the $\theta$-axes determining the critical percolation threshold for each $k$.

As it is well-known, the transition is never sharp for finite systems. Accordingly, the intersection point in previous figure is not an unique point and shows a slight shift with changes of the lattice size $L$. As we will show next, the scaling theory offers a more accurate route to determinate the percolation thresholds.

We will start by analyzing the correlation length, $\xi$. This quantity, associated with emergence of the percolating cluster, has the scaling relation:
\begin{equation}\label{ecu2}
   \xi\propto\left|\theta-\theta_{p,k} \right|^{-\nu},
\end{equation}
where $\nu$ is the critical exponent. As $\theta \to \theta^{X}_{p,k}(L)$ the correlation length $\xi \to L$, being $L$ the linear dimension of the system. Thus, we have
\begin{equation}
\label{ecu3}
   \theta^{X}_{p,k}(L)= \theta_{p,k}(\infty)+A^{X}L^{-\frac{1}{\nu}},
\end{equation}
where $A^X$ is a non-universal constant and $\theta_{p,k}(\infty)$ represents the percolation threshold in the thermodynamic limit.

As it can be seen from Eq. (\ref{ecu3}), the exponent $\nu$ is of importance because it is necessary in order to calculate the percolation threshold. The finite-size scaling theory allows to estimate $\nu$ through the scaling relationship for $R^{X}_{L,k}(p)$:
\begin{equation}\label{ecu4}
   R^{X}_{L,k}(\theta)=\overline{R^{X}_{k}}\left[ \left (\theta-\theta_{p,k}\right)L^{\frac{1}{\nu}}\right],
\end{equation}
being $\overline {R^{X}_{k}(u)}$ the scaling function and $u \equiv(\theta-\theta_{p,k})L^{\frac{1}{\nu}}$. Thus, the maximum of the derivative of Eq. (\ref{ecu4}) leads to
\begin{equation}
\left( \frac{dR^{X}_{L,k}}{d\theta} \right)_{max} \propto L^{\frac{1}{\nu}}. \label{expnu}
\end{equation}
In Fig. \ref{fig8}, this relation has been plotted as a function of $L/k$ (in log-log scale) for $k=2$. As can be observed, the slope of the curve ($1/\nu$) remains constant and close to $1.13$. Thus, the resulting value of the critical exponent was $\nu=0.91(6)$.

Another alternative way to obtain  $\nu$ is given by the divergence of the root mean square deviation of the threshold observed from their average values, $\Delta_{L,k}^A$ in Eq. (\ref{gauss})
\cite{Stauffer},
\begin{equation}
\Delta_{L,k}^X \propto L^{-1/\nu}. \label{delta}
\end{equation}
The inset of Fig. \ref{fig8} shows $\ln \left(\Delta_{L,k}^A \right)$ as a function of $\ln(L/k)$ (note the log-log functional dependence) for $k=2$. According to Eq. (\ref{delta}), the slope corresponds to $-1/ \nu$. In this case, it results $\nu=0.92(8)$.

The study in Fig. \ref{fig8} was repeated for the $I$ and $U$ percolation criteria and different values of $k$ ranging between 2 and 16. In all cases, the results coincide, within numerical errors, with the well-known value of the critical exponent of the ordinary 3D percolation $\nu=0.8774(13)$ \cite{Koza}.

\begin{figure}[h!]
	\centering
	\includegraphics[width=10cm,clip=true]{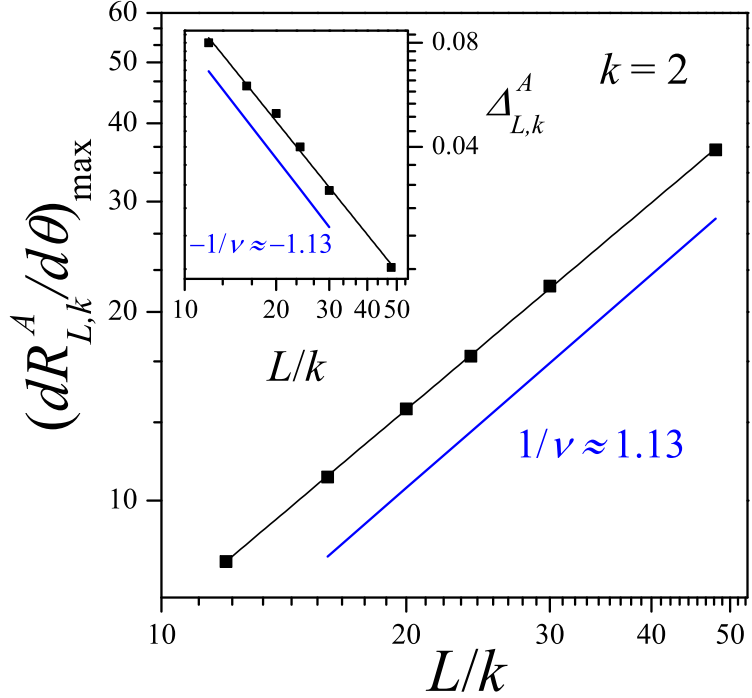}
	\caption{$\ln[(dR_{L,k}^A/d\theta)_{max}]$ as a function of $\ln(L/k)$. According to Eq. (\ref{ecu4}) the slope corresponds to $1/\nu$. Inset: $\ln(\Delta_{L,k}^A)$ versus $\ln(L/k)$ from Eq. (\ref{gauss2}). The corresponding slope should be $-1/\nu$. The data in the figure were obtained for the case $k=2$. }
	\label{fig8}
\end{figure}

The values of $\theta^X_{p,k}(L)$ can be obtained for different values of $k$ and $L$ by fitting the corresponding $dR^{X}_{L,k}(\theta)/d\theta$ curves according to Eq. (\ref{gauss2}). Then, once $\nu$ was determined, the percolation thresholds $\theta^X_{p,k}(\infty)$ can be calculated by using Eq. (\ref{ecu3}). This procedure is shown in Fig. \ref{fig6} for a typical case: $k=2$. The figure supports the relation given by Eq. (\ref{ecu3}): (a) all the curves (different criteria) are well correlated by a linear function, and (b) they have a quite similar value for the ordinate in the limit $L \to \infty$.

\begin{figure}[h!]
	\centering
	\includegraphics[width=12cm,clip=true]{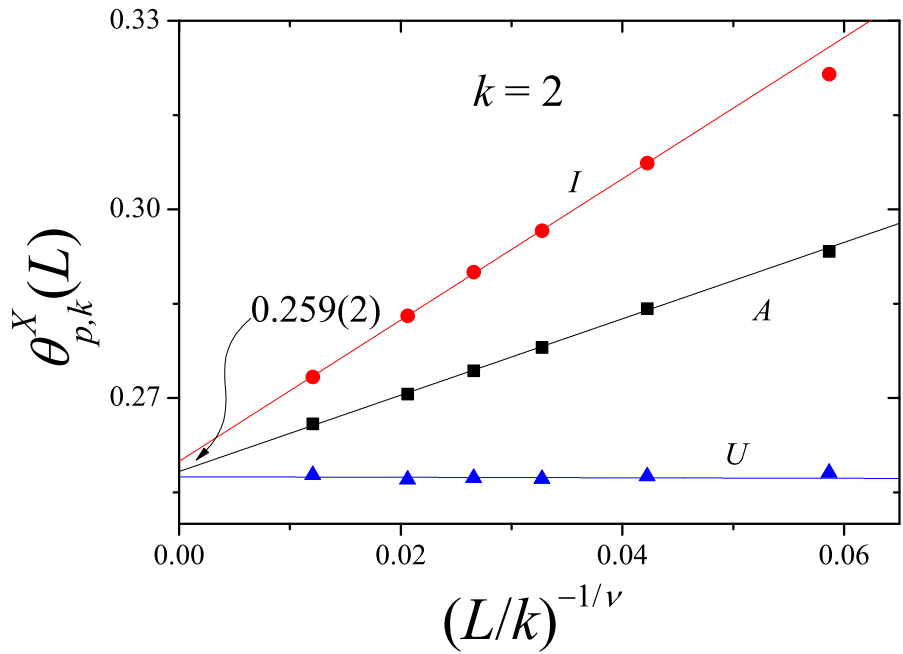}
	\caption{Extrapolation of $\theta$ towards the thermodynamic limit according to the theoretical prediction given by Eq. (\ref{ecu3}) for $k=2$. Circles, squares and triangles correspond to the criteria $I$, $A$ and $U$, respectively.}
	\label{fig6}
\end{figure}

From the procedure shown in Fig. \ref{fig6}, one obtains $\theta^X_{p,k}(\infty)$ for the criteria $I$, $A$ and $U$. Combining the three estimates for each $k$, the final values of $\theta_{p,k}(\infty)$ are obtained. The maximum of the differences between $\theta^I_{p,k}(\infty)$, $\theta^A_{p,k}(\infty)$, and $\theta^U_{p,k}(\infty)$, gives the error bar for $\theta_{p,k}(\infty)$. In the case of Fig. \ref{fig6}, the value obtained for the percolation threshold was: $\theta_{p,k=2}(\infty)=0.259(2)$. For the rest of the paper, we will denote the percolation threshold for each size $k$ by $\theta_{p,k}$ [for simplicity we will drop the symbol``$(\infty)$"].

In Fig. \ref{fig7} the percolation threshold $\theta_{p,k}$ is plotted as a function of $k$ (open squares). The corresponding numerical values are collected in Table I (third column). The figure also includes the jamming curve $\theta_{j,k}$ (solid squares). For $2 \leq k \leq 16$, the percolation threshold increases upon increasing $k$. For $k > 16$, all jammed configurations are non-percolating states, and consequently, there is no percolating phase transition. Jamming and percolation can simultaneously occur in these systems up to $k_{max}=16$, with diminishing values for the jamming critical coverage. For larger values of $k$, the jamming critical concentration happens earlier than the likely percolation concentration thus suppressing this property. This phenomenon can be better visualized in the inset of Fig. \ref{fig7}, where the ratio $\theta_{p,k}/\theta_{j,k}$ has been plotted as a function of $k$. A similar behavior has already been observed in a system of $k \times k$ tiles on square lattices, being in this case $k_{max}=3$ \cite{Nakamura86,Nakamura87,Centres2018}.

On the other hand, the result shown in Fig. \ref{fig7} contrasts with the behavior observed in systems of $n_o$-dimensional objects deposited on $n_L$-dimensional lattices (with $n_o<n_L$): straight rigid $k$-mers on 2D square lattices \cite{Becklehimer,Vandewalle,Corne1,Corne2,Leroyer,Bonnier,Kondrat,Tara2012,Kondrat2017,Slutskii}; straight rigid $k$-mers on 3D simple cubic lattices \cite{EPJB4} and $k \times k$ tiles ($k^2$-mers) on 3D simple cubic lattices \cite{Baldosas3D}. In these systems, percolating and non-percolating phases extend to infinity in the parameter space $k$ and, consequently, the model presents percolation transition in all ranges of $k$-mer size.

\begin{figure}[h!]
	\centering
	\includegraphics[width=12cm,clip=true]{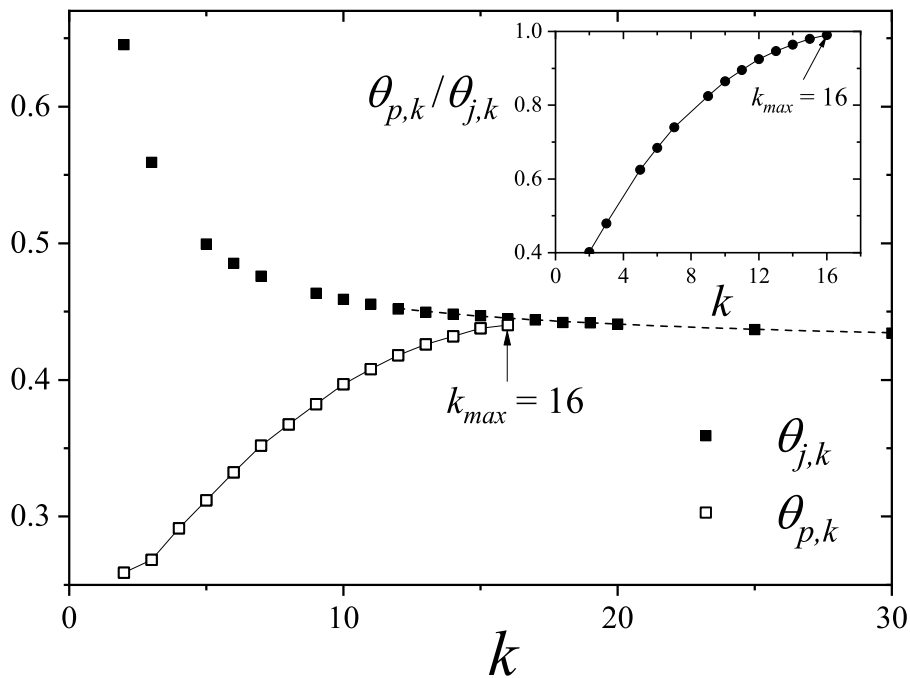}
	\caption{The thresholds $\theta_{j,k}$ (black squares) and $\theta_{p,k}$ (hollow squares) as a function of $k$. Inset: ratio $\theta_{p,k}/\theta_{j,k}$ as a function of $k$. The dashed line represent the best fit of the numerical values as indicated in the text.}
	\label{fig7}
\end{figure}

The values previously calculated for the critical exponent $\nu$ (see Fig. \ref{fig8}) clearly indicate that the percolation phase transition belongs to the the universality class of 3D random percolation. In order to reinforce or discard this hypothesis, the critical exponents $\beta$ and $\gamma$ can be calculated from the scaling behavior of $P$ and $\chi$ \cite{Stauffer} as follows:
\begin{equation}\label{ecu:scalingbeta}
P=L^{-\beta/\nu} \overline{P} \left[ |\theta -\theta_{p,k}| L^{1/\nu} \right],
\end{equation}
and
\begin{equation}\label{ecu:scalinggama}
\chi = L^{\gamma/\nu} \overline{\chi} \left[ (\theta - \theta_{p,k}) L^{1/\nu} \right],
\end{equation}
where $\overline{P}$ and $\overline{\chi}$ are scaling functions for the respective quantities.

Then, given $\theta_{p,k}$ and $\nu=XX$, $\beta$ and $\gamma$ were obtained by plotting $PL^{\beta/\nu}$ versus $|\theta -\theta_{p,k}| L^{1/\nu}$  and $\chi  L^{-\gamma/\nu}$ versus $(\theta - \theta_{p,k}) L^{1/\nu}$ and looking for data collapsing \cite{Stauffer}. As is shown in Fig. \ref{fig9}, the data scaled extremely well using $\beta=0.41$ and $\gamma=1.82$ \cite{Stauffer}.  Figure \ref{fig9} also includes the data collapse of $R^{A}_{L,k}(\theta)$ versus $(\theta - \theta_{p,k}) L^{1/\nu}$.

\begin{figure}[h!]
	\centering
	\includegraphics[width=10cm,clip=true]{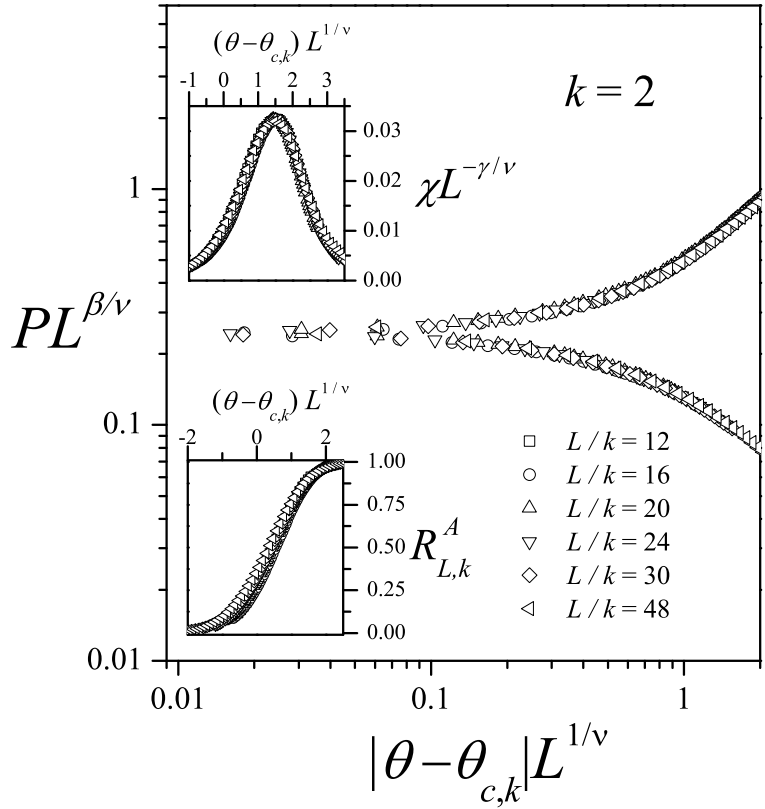}
	\caption{Data collapse of the order parameter, $PL^{\beta/\nu}$ versus $\lvert\theta-\theta_{p,k}\rvert L^{1/\nu}$ (main figure), the susceptibility $\chi L^{-\gamma/\nu}$ versus $(\theta-\theta_{p,k})L^{1/\nu}$ (upper inset), and the fraction of percolation samples $R^{A}_{L,k}(\theta)$ versus $(\theta - \theta_{p,k}) L^{1/\nu}$ (lower inset), for $k=2$. The plots were made using the percolation exponents $\nu=0.8774$, $\beta = 0.41$ and $\gamma =1.82$. }
	\label{fig9}
\end{figure}

The results obtained in Fig. \ref{fig9} support the hypothesis that the model studied here belongs to the universality class of 3D random percolation \cite{Stauffer,Koza} (see Wikipedia webpage: https://en.wikipedia.org/wiki/Percolation$_-$critical$_-$exponents). Identical results were found for different values of $k$ in the range $2 \leq k \leq 16$ (not shown here for space reasons), showing that the universality does not depend on the $k^3$-mer size. This kind of behavior, which is expected for systems without long-range correlations, has been observed in previous studies of percolation of extended objects. Thus, Cornette et al \cite{Corne2} found that straight rigid $k$-mers and tortuous $k$-mers isotropically deposited on two-dimensional square lattices are in the same universality class as the 2D standard percolation. The same result was obtained for percolation of aligned rigid rods \cite{Longone} and percolation of rigid rods under equilibrium conditions \cite{Matoz} on 2D square lattices. In the case of 3D systems, Garc\'{\i}a et al \cite{PHYSA38} arrived at the same conclusion studying the RSA process of rods. The authors reported that even though the intersection points of the curves of $R^{X}_{L,k}(\theta)$ for different objects sizes exhibit nonuniversal critical behavior, the percolation transition occurring in the system belongs to the standard 3D random percolation universality class regardless of the value of $k$ considered.

\section{Conclusions} \label{conclu}

Jamming and percolation properties in RSA of $k \times k \times k $ cubic objects ($k^3$-mers) deposited on simple cubic lattices have been studied by numerical simulations complemented with finite-size scaling theory.

The dependence of the jamming coverage $\theta_{j,k}$ on the size $k$ was studied for $k$ ranging from 2 to 40. A decreasing behavior was observed for $\theta_{j,k}$, with a finite value of saturation in the limit of infinitely long $k^3$-mers: $\theta_{j,k}= A + B/k + C/k^2$ $(k \geq 12)$, being $A=\theta_{j,k=\infty}$=0.4204(9), $B$=0.44(3) and $C$=-0.75(30). A similar decreasing behavior was found for RSA of linear $k$-mers \cite{PHYSA38} and $k \times k$ tiles \cite{Baldosas3D} on simple cubic lattices. However, some important differences between these systems were observed: $(1)$ in the range studied by simulations ($2 \leq k \leq 40$), the $k^3$-mers are less effective in filling the 3D cubic lattice than other less compact objects (such as linear $k$-mers and $k \times k$ tiles); and $(2)$ the tendency described in point $(1)$ seems to become invalid for large values of $k$, and $\theta_{j,k=\infty}$=0.4045(19) \cite{PHYSA38}, 0.4285(6) \cite{Baldosas3D} and 0.4204(9), for $k$-mers, $k^2$-mers and $k^3$-mers, respectively. Conclusion $(2)$ is based on extrapolating simulation results obtained for relatively small $k$ sizes. Accordingly, more simulations are necessary in order to obtain direct confirmation of these predictions.

To conclude with the analysis of jamming properties, the critical exponent $\nu_j$ was measured for different values of the size $k$. In all cases, the values obtained for $\nu_j$ (1) remain close to 3/2, (2) coincide, within numerical errors, with the same value of the critical exponent obtained by us for other three dimensional systems \cite{PHYSA38,Baldosas3D}, and (3) differs clearly from the value $\nu_j \approx 1$ reported by Vandewalle et al. \cite{Vandewalle} for the case of linear $k$-mers on square lattices, and from other 2D systems \cite{Centres2018,JSTAT9}.

Once the limiting parameters $\theta_{j,k}$ were determined, the percolation properties of the system were studied. The numerical calculations showed that the percolation threshold is an increasing function of $k$ in the range $2 \leq k \leq 16$. For $k \geq 17$, all jammed configurations are non-percolating states, and consequently, the percolation phase transition disappears. The interplay between the percolation and the jamming effects is responsible for the existence of a maximum value of $k$ (in this case, $k = 16$) from which the percolation phase transition no longer occurs. A similar behavior was observed in the case of $k \times k$ square tiles on 2D square lattices, where the percolation phase transition disappears for $k \geq 4$ \cite{Nakamura86,Nakamura87,Centres2018}.

Finally, and in order to test the universality of the problem, the phase transition involved on it has been studied by using finite-size scaling theory. The accurate determination of critical exponents ($\nu$, $\gamma$ and $\beta$) revealed that the model belongs to the same universality class as the 3D random percolation, regardless of the size $k$ considered. In addition, the corresponding curves collapse according to the predictions of the scaling theory.

The results obtained in the present study, along with the data reported by us and others previously \cite{Stauffer,Evans,Nakamura86,Nakamura87,Centres2018,Tara2012,Baldosas3D,Kondrat2017,Slutskii,EPJB4}, allow us to state the following classification, according to the relationship between the dimension of the depositing object and the dimension of the substrate:
\begin{itemize}
\item{$D$-dimensional lattice and $D$-dimensional depositing object:} The percolation threshold is an increasing function of $k$ in the range $2 \leq k \leq k_{max}$. For $k > k_{max}$, all jammed configurations are non-percolating states, and consequently, the percolation phase transition disappears. Thus, (1) $k_{max}=1$ for straight rigid $k$-mers on 1D lattices \cite{Stauffer,Evans}; (2) $k_{max}=3$ for $k \times k$ square tiles ($k^2$-mers) on 2D square lattices \cite{Nakamura86,Nakamura87,Centres2018}; and (3) $k_{max}=16$ for $k \times k \times k $ cubic objects ($k^3$-mers) deposited on 3D simple cubic lattices (this work).

\item{$D$-dimensional lattice and ($D-1$)-dimensional depositing object:} The percolation threshold is a nonmonotonic function of the size $k$: it decreases for small particle sizes, goes through a minimum around $k = k_{min}$, and finally tends to a constant value for large $k$'s. In other words, the percolation phase transition occurs for all values of $k$. Thus, $k_{min}=13$ for straight rigid $k$-mers on 2D square lattices \cite{Tara2012} and $k_{min}=18$ for $k^2$-mers on 3D simple cubic lattices \cite{Baldosas3D}. The tendency to a constant value for large objects has been established only for straight rigid $k$-mers on square lattices. In this line, Kondrat et al. \cite{Kondrat2017} presented a strict proof that in any jammed configuration, all clusters are percolating clusters. Later, the results obtained in Ref. \cite{Kondrat2017} were corroborated using simulation techniques \cite{Slutskii}. In the case of $k^2$-mers on 3D simple cubic lattices, direct simulation of the RSA
for such large objects is a very-time-consuming task and, therefore, it is still an open problem.

\item{$D$-dimensional lattice and ($D-2$)-dimensional depositing object:} This case corresponds to straight rigid $k$-mers on 3D simple cubic lattices. The percolation threshold shows a monotonic decrease with the size $k$ and remains below the curve of jamming coverage versus $k$. Consequently, percolating and non-percolating phases extend to infinity in the space of the parameter $k$ and the model presents percolation transition in all ranges of said value \cite{EPJB4}.
\end{itemize}

\section{ACKNOWLEDGMENTS}

This work was supported in part by CONICET (Argentina) under project number PIP 112-201101-00615; Universidad Nacional de San Luis (Argentina) under project No. 03-0816; and the National Agency of Scientific and Technological Promotion (Argentina) under project  PICT-2013-1678. The numerical work were done using the BACO parallel cluster (composed by  50 PCs each with an Intel i7-3370 / 2600 processor) located  at Instituto de F\'{\i}sica Aplicada, Universidad Nacional de San Luis - CONICET, San Luis, Argentina.

\end{large}

\end{document}